\documentclass{article}
\usepackage[margin=1in]{geometry}
\usepackage{amsmath}
\usepackage{epsfig}
\numberwithin{equation}{section} 

\newcommand{\beq}{\begin{equation}}
\newcommand{\eeq}{\end{equation}}
\newcommand{\beqa}{\begin{eqnarray}}
\newcommand{\eeqa}{\end{eqnarray}}
\newcommand{\bdm}{\begin{displaymath}}
\newcommand{\edm}{\end{displaymath}}
\newcommand{\lslash}[1]{#1\llap/}

\newcommand{\Eq}[1]{Eq.\ (\ref{#1})}
\newcommand{\Eqs}[2]{Eqs.\ (\ref{#1}) and (\ref{#2})}

\newcommand{\Rref}[1]{Ref.\ \cite{#1}}

\newcommand{\Tr}{\mbox{Tr}\,}
\newcommand{\Fig}[1]{Fig.\ \ref{#1}}

\newcommand{\Section}[1]{Section\ \ref{#1}}
\newcommand{\Appendix}[1]{Appendix\ \ref{#1}}
\newcommand{\bra}[1]{\langle #1|}
\newcommand{\ket}[1]{|#1\rangle}
\newcommand{\braket}[2]{\langle #1|#2\rangle}
\newcommand{\omegap}{\omega^{(+)}}
\newcommand{\omegam}{\omega^{(-)}}

\newcommand{\vkappa}{\vec\kappa}

\newcommand{\vecP}{\vec P}

\title{
  Fermion scattering in a Bose-Einstein condensate
}
\date{March 2026}
\author{
  C\'esar E. Echevarr\'{\i}a$^a$\footnote{cesar.echevarria@upr.edu},
  Jos\'e F. Nieves$^a$\footnote{nieves@ltp.uprrp.edu},
  Francisco Orbe$^a$\footnote{francisco.orbe@upr.edu},
  Sarira Sahu$^b$\footnote{sarira@nucleares.unam.mx}\\[12pt]
  ${}^{(a)}$Laboratory of Theoretical Physics, Department of Physics\\
  University of Puerto Rico, R\'{\i}o Piedras, Puerto Rico 00936
  \\[12pt]
  ${}^{(b)}$Instituto de Ciencias Nucleares\\
  Universidad Nacional Aut\'onoma de Mexico\\
  Circuito Exterior, C. U.\\
  A. Postal 70-543, 04510 Mexico DF, Mexico\\
}

\begin{document}
\maketitle

\begin{abstract}
  In a recent work, we considered the propagation of fermions
  in the background of a scalar Bose-Einstein (BE) condensate.
  Using some illustrative Yukawa-type coupling models between the
  fermions and the scalar fields, we determined the dispersion relations
  of the fermions and the scalar modes in various models.
  To complement that work, here we consider the corresponding fermion spinors
  and propagators, which are required for the calculation of rates
  of processes involving the fermions, as well as the thermal, and/or
  higher order, corrections to such rates. We obtain and present here
  concise formulas which are useful for those applications.
  As an application and for illustrative purposes we consider the scattering
  rate for a generic fermion (which we denote by $\chi$) with the
  fermion modes ($f$) in the BE condensate, using commonly used models for
  the $f\chi$ interactions.
  Due to the fact that the fermion modes dispersion
  relations are helicity dependent, the kinematics have
  some unique and non-standard features. For example,
  at a particular value of the momentum one of the fermion modes
  has zero group velocity, which leads to effects of the type of
  the Van Hove singularity in the density of states of
  some condensed matter systems. The results of the framework presented here,
  besides their merit in their own right, can be useful in specific contexts
  and applications, such as cosmic-ray electron cooling through
  dark matter-electron scattering, and similar ones involving
  neutrino and/or electron propagation in a Dark Matter background. 
\end{abstract}

\section{Introduction and summary}
\label{sec:intro}

It is well known that the dispersion relations of neutrinos that
propagate in a medium are modified in important ways due to the interactions
of the neutrinos with the particles in the background.
Besides the effects due the standard weak interactions when
the neutrinos propagate through normal matter,
many works have considered the effects due to non-standard neutrino
interactions and/or backgrounds in a variety of physical contexts. 
A significant effort along those lines involve extensions of
the standard electroweak model in which the neutrinos interact
with a scalar ($\phi$) and fermion ($f$) by a Yukawa-type interaction of the
form $\bar f_R\nu_L\phi$\cite{babu,ns:nuphiresonance}. The effects when
the neutrinos propagate in a medium that contains a background of those
particles have been considered in a variety
of physical contexts, including collective oscillations in supernova
\cite{Duan:2010bg,Chakraborty:2016yeg,Wong:2002fa},
the hot plasma of the early universe \cite{PhysRevD.72.085016}, cosmological
observations such as cosmic microwave background \cite{Sellentin_2015},
Big Bang nucleosynthesis data \cite{Steigman_2012,Mangano:2006ar}, and in
particular dark matter-neutrino interactions \cite{%
  Mangano:2006mp,Binder:2016pnr,Primulando:2017kxf,
  Campo:2017nwh,Pandey:2018wvh}.
 
Motivated by these developments, we have considered in previous
works the calculations of various effects, including the damping
and decoherence effects, on the neutrino propagation in such
scalar and fermion backgrounds\cite{ns:nuphiresonance}. Of particular relevance
to the present work, in \Rref{ns:fbec} we presented
a method to determine the effective potential and dispersion relation
of a fermion in the background of the BE condensate,
in a general way and not tied to any specific application.
The idea that the dark matter (DM) can be self-interacting
is intriguing, and a DM background of scalar particles is a
candidate for such environments
\cite{garani:becdm,kirkpatrick:becdm,bohmen:becdm,craciun:becdm}.
In that context, our interest is the application to the case of a neutrino
propagating in such backgrounds.

Besides the modification of the dispersion relation, the interaction
also modifies the wavefunction associated with the propagating mode.
The determination of the wavefunction is required to calculate in a
consistent way the rates of processes involving the propagating particle.
For example, in the case of fermions, the correct spinor to use
is the solution to the field equation corresponding to the dispersion
relation of the propagating mode\cite{jcdjfn:fermdamp}.

In this work we complement \Rref{ns:fbec}, by considering the determination
of the fermion spinors in the models considered there.
To be specific, here we obtain and present
concise formulas for the spinor projection matrices which are useful for
calculations of process rates involving a fermion propagating in a 
BE condensate. As an application and for illustrative purposes we specifically
consider the rate for the scattering process of
generic fermion (which we denote by $\chi$) with the
fermions in the BE condensate. The results presented here, besides
their merit in their own right and serving as a guide for similar models,
can be useful in specific contexts and applications
that involve neutrino, electron or nucleon interactions
with dark matter particles\cite{ghosh1,guha,cappiello,ghosh2}.

For guidance and reference, in \Section{sec-preliminaries} we borrow
material from \Rref{ns:fbec} to briefly
review the method we use for treating fermions $f$ propagating
in a Bose-Einstein (BE) condensate, in particular
the fermion dispersion relations in the model (Model-I of that reference)
that we use for the example calculations in the present work.
In \Section{sec-spinors} we determine the spinors that follow
from the field equations, corresponding to the dispersion relations.
More importantly for our purposes here, we determine and present the
corresponding formulas for the spinor projection matrices,
in a convenient form suitable for the type of calculations
we are considering, namely transition rates involving the fermion modes.
In \Section{sec-propagator} we obtain the formulas
for the corresponding propagators, including their thermal
counterpart. Although the formulas are not used directly in the present
work, they are nevertheless required in order to determine
the proper normalization factors of the spinors and the projection
matrices obtained in \Section{sec-spinors}.  

As example application of the fermion spinors and projection matrices,
in \Section{sec-scatt} we consider the calculation of rates for processes
involving the interaction of a fermion $\chi$ with a fermion mode $f$.
For definiteness, we consider in detail the situation in which
$\chi$ is a heavy fermion, which can be considered at rest. As simple
as this case may seem at first sight, the analysis and results illustrate
several technicalities that must be dealt with in this type of calculation
involving the $f$ fermions, that are due to the fact that the
dispersion relations of the $f$ modes are not
the standard vacuum dispersion relations, but in fact
are helicity-dependent, and therefore the kinematics
differ from the case in which the fermions propagate in vacuum.
For example, among other things, a particular consequence is the fact that
there is a kinematical range in which the group velocity
of one of the $f$ modes decreases as the momentum increases,
and in fact vanishes at a particular point,
reminiscent of the Van Hove singularity in Condensed Matter
systems\cite{vanhovesing,vanhovenature}.

We give our conclusions and outlook in \Section{sec-conclusions},
and in \Appendix{sec-appendixA} we give a compendium of the relevant
formulas for the fermion dispersion relations, spinors, projection operators
and propagators that are useful for reference in future works.

\section{Preliminaries}
\label{sec-preliminaries}

For reference, and to state clearly the context of our work, we present
in this section the model we use to treat the fermions and the BE condensate
and summarize the main results obtained in \Rref{ns:fbec} that we need
in the present work.

\subsection{Model}
\label{sec:modeli}

We consider the model labeled as ``Model-I'' in \Rref{ns:fbec}. It consists
of a complex scalar field $\phi$, and two chiral fermions $f_L$ and $f_R$,
with an interaction Lagrangian
\beq
\label{Lfphi}
L_{\text{int}} = -\lambda \phi \bar f_R f_L + h.c\,.
\eeq
Apart from the interaction term given above, the Lagrangian
has the fermion terms
\beq
L_{f} = i\bar f_L\lslash{\partial} f_L +
i\bar f_R\lslash{\partial}f_R\,,
\eeq
and a standard $\phi^4$ Lagrangian
\beq
\label{Lphi}
L^{(\phi)} =  (\partial^\mu\phi)^\ast(\partial_\mu\phi) - V_0\,,
\eeq
with
\beq
\label{V0}
V_0(\phi) = m^2_\phi\phi^\ast\phi + \lambda_\phi(\phi^\ast\phi)^2\,,
\qquad (\lambda_\phi > 0)\,.
\eeq
There are two conserved charges, labeled as $Q_{1,2}$, with
\beqa
Q_1(f_L) = Q_1(f_R) = 1,& \qquad & Q_1(\phi) = 0\,,\nonumber\\
Q_2(\phi) = Q_2(f_R) = 1, & \qquad & Q_2(f_L) = 0\,,
\eeqa
which satisfy
\beq
\label{chargeassignments}
Q_A(\phi) + Q_A(f_L) - Q_A(f_R) = 0\,.
\eeq
The thermal expectation value of any operator $O$ is given by
\beq
\langle O\rangle = \frac{\Tr ZO}{\Tr Z}\,,
\eeq
where $Z$ is the partition function operator
\beq
\label{Zgen}
Z = e^{-\beta H + \sum_A\alpha_A Q_A}\,.
\eeq
The assignments in \Eq{chargeassignments} imply that the chemical
potentials satisfy
\beq
\label{murelation}
\mu_\phi + \mu_L - \mu_R = 0\,,
\eeq
where we are denoting by $\mu_L$ and $\mu_R$ the
chemical potential of $f_L$ and $f_R$, respectively.

The method used in \Rref{ns:fbec} consists in rewriting the Lagrangian
in terms of the fields $\phi^\prime, f^\prime_L, f^\prime_R$, defined by
\beqa
\label{primedfieldsi}
\phi & = & e^{-i\mu_\phi t}\phi^\prime\,,\nonumber\\
f_L & = & e^{-i\mu_L t}f^\prime_L\,,\nonumber\\
f_R & = & e^{-i\mu_R t}f^\prime_R\,.
\eeqa
As explained in \Rref{ns:fbec},
we can then use the partition function operator
without the chemical potential terms
$\sum_A\alpha_A Q_A$, provided we use the Lagrangian
written in terms of the prime fields $\phi^\prime, f^\prime_L, f^\prime_R$.
This idea is motivated by the results of previous works by other
authors\cite{footnote1} that consider the BE condensation of scalar fields.
However, we emphasize that the proof given in Appendix A of \Rref{ns:fbec}
is a proof at the operator level not involving Lagrangian dynamics arguments,
which uses only the equation of motion and the commutators of the charges
with the field, it does not assume that it is a free field,
and it is applicable to any field (e.g., a fermion field).
Therefore the result holds for any field, not just a scalar field,
whether it is interacting or not.

For our current purposes the upshot is that the prime fields are the
appropriate ones that describe the modes in the BE condensate, in particular
$f^\prime_R$ and $f^\prime_L$ for the fermion modes.
For practical purposes, the recipe is to substitute the transformations
given in \Eq{primedfieldsi} into the Lagrangian to obtain
the Lagrangian in terms of the prime fields. The interaction term
in \Eq{Lfphi} does not change,
\beq
\label{Lfphiprime}
L_{\text{int}} = -\lambda \phi^\prime \bar f^\prime_R f^\prime_L + h.c\,.
\eeq
Regarding the scalar field, the result of substituting
$\phi = e^{-i\mu_\phi t}\phi^\prime$ in $L^{(\phi)}$, which we denote
by $L^{(\phi^\prime)}$, is
\beq
\label{Lphiprime}
L^{(\phi^\prime)} = K - U\,,
\eeq
where
\beq
\label{U}
U = -(\mu^2_\phi - m^2_\phi)\phi^{\prime\,\ast}\phi^\prime +
\lambda_\phi(\phi^{\prime\,\ast}\phi^\prime)^2\,,
\eeq
and $K$ contains the terms with derivatives of $\phi^\prime$. We do not
write them explicitly because we will not need them in what follows,
but they are of course important to determine the properties (e.g.,
dispersion relations) of the scalar modes, and are discussed at length
in \Rref{ns:fbec}.

Therefore, if the conditions are such that
\beq
\label{sbcondition}
\mu^2_\phi > m^2_\phi\,,
\eeq
the minimum of the potential $U$ is not at $\phi^\prime = 0$,
and therefore $\phi^\prime$ develops a non-zero expectation
value. The $U(1)$ symmetry corresponding $Q_2$ is broken,
which is associated with BE transition. Explicitly,
\beq
\label{phiexpectationvalue}
\langle\phi^\prime\rangle \equiv \frac{1}{\sqrt{2}}\phi_0\,,
\eeq
\beq
\label{phi0}
\phi^2_0 = \frac{\mu^2_\phi - m^2_\phi}{\lambda_\phi}\,.
\eeq
As in \Rref{ns:fbec}, we assume that this condition is satisfied.

Regarding the fermion fields, as a result of making the substitution
given in \Eq{primedfieldsi}, together with the symmetry breaking effect
just described, the bilinear fermion terms in the Lagrangian are
\beq
L_0 = \bar f^\prime_L i\lslash{\partial} f^\prime_L + 
\bar f^\prime_R i\lslash{\partial} f^\prime_R +
\mu_L\bar f^\prime_L\lslash{u} f^\prime_L +
\mu_R \bar f^\prime_R \lslash{u} f^\prime_R -
(m\bar f^\prime_R f^\prime_L + h.c.)\,,
\eeq
where
\beqa
\label{msymmbreaking}
m & = & \frac{\lambda\phi_0}{\sqrt{2}}\,,\nonumber\\
& = & \frac{\lambda}{\sqrt{2}}\left(
\frac{\mu^2_\phi - m^2_\phi}{\lambda_\phi}\right)^{1/2}\,,
\eeqa
and $u^\mu$ is the four-vector
\beq
\label{u}
u^\mu = (1,\vec 0)\,.
\eeq
The complete Lagrangian, in terms of the prime fields, is
\beq
\label{modelitotalL}
L = L_0 + L^{(\phi^\prime)} + L^\prime_{\text{int}}\,,
\eeq
where $L^{(\phi^\prime)}$ is given in \Eq{Lphiprime}, while
$L^\prime_{\text{int}}$ is the interaction term between the fermion and
the scalar fields, which we will not need in what follows.
Defining
\beq
f = f^\prime_L + f^\prime_R\,,
\eeq
in momentum space $L_0$ is given by
\beq
\label{L0k}
L_0(k) = \bar f(k)(\lslash{k} - \Sigma(k))f(k)\,,
\eeq
where
\beq
\label{Sigmai}
\Sigma = mL + m^\ast R - \mu_L\lslash{u}L - \mu_R\lslash{u}R \,.
\eeq
In this model, the two chiral fermions form a Dirac particle,
in which the left and right components have different dispersion relations.

Despite the similarity with the standard Higgs mechanism,
in the type of model we are considering the fermion mass term is induced
by the condensate, which in turn is induced by the chemical potentials.
In particular, if the chemical potentials are zero, then
\Eq{sbcondition} is not satisfied, and from \Eqs{phi0}{msymmbreaking},
$\phi_0$ and $m$ remain zero. Moreover, the chemical potentials
also generate the terms proportional to $\lslash{u}$ in \Eq{Sigmai},
reminiscent of the self-energy of a fermion in a thermal background
(e.g., a neutrino in a matter background).

\subsection{Dispersion relations}
\label{sec-dispersionrelations}

The field equation in momentum space is
\beq
(\lslash{k} - \Sigma)f = 0\,.
\eeq
We use systematically the Weyl (\emph{chiral}) representation of the
gamma matrices,
\beq
\gamma^0 = \left(\begin{array}{cc}
0 & 1\\
1 & 0
\end{array}\right) \qquad
\gamma^i = \left(\begin{array}{cc}
0 & -\sigma^i\\
\sigma^i & 0
\end{array}\right) \qquad
\gamma^5 \equiv i\gamma^0\gamma^1\gamma^2\gamma^3 = \left(\begin{array}{cc}
1 & 0\\
0 & -1
\end{array}\right)\,.
\eeq
For completeness, we also mention that we use the convention
\beq
\epsilon^{0123} = +1\,.
\eeq

Using the Weyl representation of the gamma matrices we put
\beqa
f^\prime_L & = & \left(\begin{array}{cc}0\\ \eta\end{array}\right)\,,
\nonumber\\
f^\prime_R & = & \left(\begin{array}{cc}\xi\\ 0\end{array}\right)\,.
\eeqa
Writing
\beq
\label{kmuomegakappa}
k^\mu = (\omega,\vec\kappa)\,,
\eeq
the equations to be solved then become
\beqa
\label{modelixietaeqs}
\left(\omega + \mu_L + \vec{\sigma}\cdot\vec\kappa\right)\eta -
m^\ast\xi & = & 0\,,
\nonumber\\
\left(\omega + \mu_R - \vec{\sigma}\cdot\vec\kappa\right)\xi -
m\eta & = & 0\,.
\eeqa
In the one-generation case we are considering the
phase of $m$ is irrelevant, since it can be absorbed by a field redefinition,
so that we could take $m^\ast = m$. However, since in more general
cases such field redefinitions cannot be done independently,
we keep $m$ arbitrary.

In general, leaving out the case that $\mu_R = \mu_L$
(i.e., assuming $\mu_\phi \not= 0$),
these equations have non-trivial solutions only if $\xi$ and $\eta$ are
proportional to the same eigenvector of $\vec{\sigma}\cdot\vec\kappa$;
and in any case we can choose the spinors $\xi$ and $\eta$ to be such.
Therefore, we write the solution in the form (NOTE: our convention for
$x,y$ here are opposite to the one in \Rref{ns:fbec})
\beqa
\label{xietachimodeli}
\xi & = & x\chi_s(\vkappa)\,,\nonumber\\
\eta & = & y\chi_s(\vkappa)\,,
\eeqa
where $\chi_s(\vkappa)$ is the spinor with definite helicity, defined by
\beq
\label{helicityspinor}
\left(\vec\sigma\cdot\hat\kappa\right)\chi_s(\vkappa) = 
s\chi_s(\vkappa)\,,
\eeq
with $s = \pm 1$. For a given helicity $s$, the equations for $x$ and $y$ are
\beqa
\label{xymodeli}
(\omega + s\kappa + \mu_L)y - m^\ast x & = & 0\,,\nonumber\\
(\omega - s\kappa + \mu_R)x - my & = & 0\,,
\eeqa
which imply that $\omega$ must satisfy
\beq
\label{dreq}
(\omega - s\kappa + \mu_R)(\omega + s\kappa + \mu_L) - |m|^2 = 0\,.
\eeq
Expressing $\mu_R$ and $\mu_L$ in terms of their sum and their difference,
\beq
\mu_\pm \equiv \frac{1}{2}(\mu_R \pm \mu_L)\,,
\eeq
this equation can be written in the form
\beq
\left(\omega + \mu_{+}\right)^2 -
\left(\kappa - s\mu_{-}\right)^2 - |m|^2 = 0\,.
\eeq

For each $s$, we have two solutions, one with positive $\omega$ and
another with a negative $\omega$.
They correspond to the positive and negative helicity states
of the Dirac particle and its anti-particle, which are associated with the
unbroken $Q_1$. We label the two solutions for each $s$ as $\omega^{(\pm)}_{s}$.
With this notation the solutions are
\beq
\label{drmodeli}
\omega^{(\pm)}_{s}(\vec\kappa) = \pm \omega_{0s} - \mu_{+}\,,
\eeq
where
\beq
\label{omega0s}
\omega_{0s} = \left[(\kappa - s\mu_{-})^2 + |m|^2\right]^{1/2}\,.
\eeq
Denoting the particle and anti-particle dispersion relations
by $\omega_s$ and $\bar\omega_s$, respectively, they are to be identified
according to
\beqa
\label{drmodeliphysical}
\omega_s(\vec\kappa) & = & \omega^{(+)}_s(\vec\kappa)\nonumber\\
& = & \omega_{0s} - \mu_{+}\,,\nonumber\\
\bar\omega_s(\vec\kappa) & = & -\omega^{(-)}_s(-\vec\kappa)\nonumber\\
& = & \omega_{0s} + \mu_{+}\,.
\eeqa
For future reference we record here the formula for the
group velocity,
\beq
\label{vg}
v_{gs} = \frac{\kappa - s\mu_{-}}{\omega_{0s}}\,.
\eeq

The dispersion relations were obtained in \Rref{ns:fbec}.
Our main purpose in the present work is to find the corresponding
spinor wave functions and propagator of the fermion modes,
which are required to calculate rates for processes involving the fermions
as well as thermal corrections to the dispersion relations.

Before entering into the details we would like to mention
the following. As we have mentioned in the discussion
of \Eq{primedfieldsi}, the chemical
potentials do not appear explicitly in the partition function
operator as a result of the field transformations. However,
the chemical potentials enter in the calculation of any physical
quantity through the dispersion relations, spinors and propagators.
Moreover, if we denote by $n_s$ the momentum
distribution functions of the fermion modes at some temperature $T = 1/\beta$,
and similarly for the antiparticles,
\beqa
\label{ns}
n_s & = & \frac{1}{e^{\beta\omega_s} + 1}\,,\nonumber\\
\bar n_s & = & \frac{1}{e^{\beta\bar\omega_s} + 1}\,,
\eeqa
the chemical potentials are present through $\omega_s$ and
$\bar\omega_s$. In the case
we are considering, we can take $\mu_{L,R}$ to be the independent
chemical potentials, with $\mu_\phi$ given by \Eq{murelation}.
Then \Eq{phi0}, together with \Eq{ns}, fix the background
composition of the system.  In particular at
$T \rightarrow 0$, some of the distribution functions are filled up
to a certain value $\kappa_F$ of the momentum, while others are zero,
depending on the relative value and sign of $\mu_{L,R}$. Which ones
are zero, and what is $\kappa_F$ for those that are not zero,
can be determined easily from the formulas for the dispersion
relations given in \Eq{drmodeliphysical}. The ones
that are not zero determine the fermion background composition of the
zero temperature system. Therefore, by means of the field transformation,
the effects of the chemical potentials are taken into account
through the dispersion relations we have discussed
above, and the wavefunctions and propagators that we consider next.

\section{Spinors}
\label{sec-spinors}

\subsection{Spinors for $\omegap_s$ - the $u_s$ spinors}

The $x,y$ coefficients corresponding to the $\omega^{(+)}_s$ dispersion
relation must satisfy \Eq{xymodeli}, for $\omega = \omegap_s$.
Remembering that we are identifying $\omegap_s$ with the particle dispersion
relation $\omega_s$, that is
\beqa
\label{xyomegap}
D_{Ls} y_s - m^\ast x_s & = & 0\,,\nonumber\\
D_{Rs} x_s - my_s & = & 0\,,
\eeqa
where we define
\beqa
\label{DDprime}
D_{Ls}(\vkappa) & = & \omega_s + s\kappa + \mu_L\nonumber\\
& = & \omega_{0s} + s\kappa - \mu_{-} \,,\nonumber\\
D_{Rs}(\vkappa) & = & \omega_s - s\kappa + \mu_R\nonumber\\
& = & \omega_{0s} - s\kappa + \mu_{-}\,.
\eeqa
Of course, $D_{Ls}, D_{Rs}$ satisfy
\beq
D_{Ls} D_{Rs} = |m|^2\,,
\eeq
by virtue of the dispersion relation $\omega^{(+)}_s$. It follows that
\beqa
x_s & = & z_s\sqrt{D_{Ls}}\,,\nonumber\\
y_s & = & z_s\alpha^\ast\sqrt{D_{Rs}}\,,
\eeqa
where $z_s$ is an arbitrary normalization factor, and $\alpha$ is the phase
factor of $m$, that is $\alpha = m/|m|$.

Therefore the complete Dirac spinor corresponding to $\omega^{(+)}_s$ is
\beq
u_s = z_s\left(
\begin{array}{cc}\sqrt{D_{Ls}}\\[12pt] \alpha^\ast\sqrt{D_{Rs}}
\end{array}\right)\chi_{s}\,,
\eeq
and in particular
\beq
\label{ubaru}
u_{s}\bar u_{s} = z^2_s
\left(\begin{array}{cc}
  m & D_{Ls}\\ D_{Rs} & m^\ast
\end{array}\right)P_s\,,
\eeq
where
\beq
\label{Ps}
P_s(\vkappa) \equiv \frac{1}{2}(1 + s\vec\sigma\cdot\hat\kappa)\,.
\eeq

Using the standard convention for the one-particle state 
with momentum $k^\mu_s = (\omega_s,\vkappa)$ and helicity $s$,
which we denote by $\ket{f(\vkappa,s)}$, we have
\beq
\bra{0}f(x)\ket{f(\vkappa,s)} = u_s(\vkappa) e^{-ik_s\cdot x}\,.
\eeq
The corresponding one-particle contribution to the propagator
\beq
iS_{F0}(x - y) = \braket{0|Tf(x)\bar f(y)}{0}\,,
\eeq
in momentum space [with $k^\mu \equiv (\omega,\vkappa)$], is 
\beq
\label{SF01p}
(S_{F0})_{1p}(k) = \frac{u_s(\vkappa) \bar u_s(\vkappa)}
{2\omega_s(\omega - \omega_s + i\epsilon)}\,.
\eeq
The normalization factor $z_s$ is determined by the requirement
that the one-particle contribution to the propagator given in \Eq{SF01p},
using \Eq{ubaru}, coincides with the exact propagator
near the $\omega = \omega_s$ pole. As we show in
\Section{sec-propagator} [see \Eq{1pomegap}], this condition gives
\beq
\label{zs}
z_s = \sqrt{\frac{\omega_s}{\omega_{0s}}}\,.
\eeq

We can write an expression for $u_s\bar u_s$ in a form that may be more
convenient for some calculations as follows. Remembering that we are using the
Weyl (chiral) representation of the gamma matrices, we have
\beq
-\gamma^0\gamma^5\vec\gamma =
\left(\begin{array}{cc}\vec\sigma & 0 \\ 0 & \vec\sigma\end{array}\right)\,,
\eeq
and therefore
\beq
P_s \equiv \frac{1}{2}
\left(\begin{array}{cc}
1 + s\vec\sigma\cdot\hat\kappa & 0\\
0 &  s\vec\sigma\cdot\hat\kappa
\end{array}\right) = \frac{1}{2}
(1 - s\gamma^0\gamma^5\vec\gamma\cdot\hat\kappa)\,.
\eeq
Introducing the vectors
\beqa
\label{umunmudef}
u^\mu & = & (1,\vec 0)\,,\nonumber\\
n^\mu & = & (0,\hat\kappa)\,,
\eeqa
we can then write
\beq
\label{Pscovdef}
P_s = \frac{1}{2}(1 + s\lslash{u}\gamma^5\lslash{n}) =
\frac{1}{2}(1 - s\gamma^5\lslash{u}\lslash{n})\,.
\eeq
On the other hand, we can also write
\beq
\left(\begin{array}{cc}
m & D_{Ls}\\
D_{Rs} & m^\ast
\end{array}\right) = mR + m^\ast L + D_{Ls}\gamma^0 L + D_{Rs}\gamma^0 R =
mR + m^\ast L + D_{Ls}\lslash{u} L + D_{Rs}\lslash{u} R\,,
\eeq
and with these we then have
\beq
\label{uprojcov}
u_s \bar u_s = \frac{z^2_s}{2}
(mR + m^\ast L + D_{Ls}\lslash{u} L + D_{Rs}\lslash{u} R)
(1 - s\gamma^5\lslash{u}\lslash{n})\,.
\eeq

\subsection{Spinors for $\omegam_s$ - the $v_s$ spinors}

We consider now the spinors corresponding to the $\omegam_s$
dispersion relation. The $x,y$ coefficients satisfy
\beqa
(\omegam + s\kappa + \mu_L)y_s - m^\ast x_s & = & 0\,,\nonumber\\
(\omegam - s\kappa + \mu_R)x_s - my_s & = & 0\,,
\eeqa
and putting $\omegam_s = -\bar\omega_s$ [remember \Eq{drmodeliphysical}],
\beqa
(\bar\omega_s - s\kappa - \mu_L)y_s + m^\ast x_s & = & 0\,,\nonumber\\
(\bar\omega_s + s\kappa - \mu_R)x_s + my_s & = & 0\,.
\eeqa
Defining
\beqa
\bar D_{Ls}(\vkappa) & = & \bar\omega_s - s\kappa - \mu_L\nonumber\\
& = & \omega_{0s} - s\kappa + \mu_{-}\,,\nonumber\\
\bar D_{Rs}(\vkappa) & = & \bar\omega_s + s\kappa - \mu_R\nonumber\\
& = & \omega_{0s} + s\kappa - \mu_{-}\,,
\eeqa
the equations are written in the form
\beqa
\label{xyeqomegam}
\bar D_{Ls} y_s & = & -m^\ast x_s\,,\nonumber\\
\bar D_{Rs} x_s & = & - my_s\,.
\eeqa
Notice that, since $\omegam_s$ satisfies \Eq{dreq}, $\bar\omega_s$ satisfies
\beq
\label{dreqomegabar}
(\bar\omega_s - s\kappa - \mu_L) 
(\bar\omega_s + s\kappa - \mu_R) - |m|^2 = 0\,,
\eeq
which in turn imply
\beq
\bar D_{Rs} \bar D_{Ls} = |m|^2\,.
\eeq

Therefore, the solution to \Eq{xyeqomegam} can be written as
\beqa
x_s & = & \bar z_s \sqrt{\bar D_{Ls}}\,,\nonumber\\
y_s & = & -\bar z_s \alpha^\ast \sqrt{\bar D_{Rs}}\,,
\eeqa
where $\alpha = m/|m|$ as before. The complete Dirac spinor corresponding
to $\omegam_s$, which we denote by $u^{(-)}$, is therefore
\beq
u^{(-)}_s(\vkappa) = \bar z_s\left(
\begin{array}{cc} \sqrt{\bar D_{Ls}}\\[12pt] -\alpha^\ast\sqrt{\bar D_{Rs}}
\end{array}\right)\chi_s(\vkappa)\,,
\eeq
and in particular
\beq
\label{ubarum}
u^{(-)}_s(\vkappa)\bar u^{(-)}_s(\vkappa) = \bar z^2_s
\left(\begin{array}{cc}
  -m & \bar D_{Ls}\\ \bar D_{Rs} & -m^\ast
\end{array}\right)P_s(\vkappa)\,,
\eeq
The $v$ spinor that appears in the plane wave expansion of the field is
defined as
\beqa
v_s(\vkappa) & \equiv & u^{(-)}_s(-\vkappa)\nonumber\\
& = & \bar z_s\left(
\begin{array}{cc} \sqrt{\bar D_{Ls}}\\[12pt] -\alpha^\ast\sqrt{\bar D_{Rs}}
\end{array}\right)\chi_s(-\vkappa)\,,
\eeqa
and therefore
\beq
\label{vbarv}
v_s(\vkappa)\bar v_s(\vkappa) = \bar z^2_s
\left(\begin{array}{cc}
  -m & \bar D_{Ls}\\ \bar D_{Rs} & -m^\ast
\end{array}\right)P_s(-\vkappa)\,.
\eeq
To determine the normalization factor, notice that
\beq
\label{vbarvm}
v_s(-\vkappa) \bar v_s(-\vkappa) =
u^{(-)}_s(\vkappa)\bar u^{(-)}_s(\vkappa) = \bar z^2_s
\left(\begin{array}{cc}
  -m & \bar D_{Ls}\\ \bar D_{Rs} & -m^\ast
\end{array}\right)P_s(\vkappa)\,.
\eeq

Using the standard convention for the one-(anti)particle state 
with momentum $k^\mu_s = (\bar\omega_s,\vkappa)$ and helicity $s$,
which we denote by $\ket{\bar f(\vkappa,s)}$, we have
\beq
\bra{0}\bar f(x)\ket{f(\vkappa,s)} = \bar v_s(\vkappa) e^{-ik_s\cdot x}\,,
\eeq
The corresponding one-particle contribution to the propagator, in 
momentum space [with $k^\mu \equiv (\omega,\vkappa)$], is
\beq
\label{SF01barp}
(S_{F0})_{1\bar p}(k) = \left.
\frac{v_s(\vkappa) \bar v_s(\vkappa)}
{2\bar\omega_s(\omega + \bar\omega_s - i\epsilon)}
\right|_{\vkappa \rightarrow -\vkappa}\,.
\eeq
The normalization factor $\bar z_s$ is determined by the requirement
that the one-particle contribution to the propagator given in \Eq{SF01barp},
using \Eq{vbarvm}, coincides with the exact propagator
near the $\omega = \omegam_s$ pole. As we show in
\Section{sec-propagator} [see \Eq{1pomegap}], this condition gives
\beq
\label{zsbar}
\bar z_s = \sqrt{\frac{\bar\omega_s}{\omega_{0s}}}\,.
\eeq

Similar to what we did for the $u_s\bar u_s$ projection operator,
it is easy to see that in this case the expression corresponding to
\Eq{uprojcov} is
\beq
\label{vprojcov}
v_s \bar v_s = \frac{\bar z^2_s}{2}
(-mR - m^\ast L + \bar D_{Ls}\lslash{u} L + \bar D_{Rs}\lslash{u} R)
(1 + s\gamma^5\lslash{u}\lslash{n})\,.
\eeq

\subsection{Limit $\mu_{R,L} = 0$ and $m^\ast = m$}

\subsubsection{$u\bar u$}

In this limit $z_s = 1$, and
\beqa
\label{ubaruvacuumlimit}
2u_s\bar u_s & = &
(m + \omega\lslash{u} - s\kappa\lslash{u}\gamma^5)
(1 - s\gamma^5\lslash{u}\lslash{n})\nonumber\\
& = & (m + \omega\lslash{u} - s\kappa\lslash{u}\gamma^5) +
\left[-sm\gamma^5\lslash{u}\lslash{n} + s\omega\gamma^5\lslash{n} +
  \kappa\lslash{n}\right]\nonumber\\
& = & m + \omega\lslash{u} + \kappa\lslash{n} + s\kappa\gamma^5\lslash{u}
+ s\omega\gamma^5\lslash{n} - sm\gamma^5\lslash{u}\lslash{n}\nonumber\\
& = & m + \lslash{k} +
sm\gamma^5\lslash{\ell} - sm\gamma^5\lslash{u}\lslash{n}\,,
\eeqa
where
\beq
\label{ell}
\ell^\mu = \frac{1}{m}(\kappa u^\mu + \omega n^\mu) =
\frac{1}{m}(\kappa,\omega\hat\kappa)\,.
\eeq

On the other hand, the standard expression in vacuum is
\beqa
2u_s\bar u_s & = & (\lslash{k} + m)(1 + s\gamma^5\lslash{\ell})\nonumber\\
& = & \lslash{k} + m + sm\gamma^5\lslash{\ell} -
s\gamma^5\lslash{k}\lslash{\ell}\,.
\eeqa
Substituting in $\lslash{k}\lslash{\ell}$ the relations
\beqa
k^\mu & = & \omega u^\mu + \kappa \vec n\,,\nonumber\\
\ell^\mu & = & \frac{1}{m}(\kappa u^\mu + \omega n^\mu)\,,
\eeqa
and using the identities
\beq
u^2 = 1\,,\quad n^2 = -1\,, \quad u\cdot n = 0\,,
\eeq
it follows that
\beq
\label{kellidentity}
\lslash{k}\lslash{\ell} = m\lslash{u}\lslash{n}\,,
\eeq
which proves that \Eq{ubaruvacuumlimit} coincides with the standard
expression in vacuum.

\subsubsection{$v\bar v$}

We proceed in a similar way for $v\bar v$. In this limit $\bar\omega = \omega$,
and therefore
\beqa
2v_s\bar v_s & = & (-m + \omega\lslash{u} + s\kappa\lslash{u}\gamma^5)
(1 + s\gamma^5\lslash{u}\lslash{n})\nonumber\\
& = &
-m + \omega\lslash{u} + s\kappa\lslash{u}\gamma^5 -
sm\gamma^5 \lslash{u}\lslash{n} - s\omega\gamma^5\lslash{n} + \kappa\lslash{n}
\nonumber\\
& = & -m + \lslash{k} -sm\gamma^5\lslash{\ell} -sm\gamma^5\lslash{u}\lslash{n}
\,,
\eeqa
while the standard expression in vacuum is
\beqa
2v_s\bar v_s & = & (\lslash{k} - m)(1 + s\gamma^5\lslash{\ell})\nonumber\\
& = & \lslash{k} - m -sm\gamma^5\lslash{\ell} -
s\gamma^5\lslash{k}\lslash{\ell}\,.
\eeqa
Using \Eq{kellidentity} once more verifies the equivalence.

\section{Propagator}
\label{sec-propagator}
  
The expression for the propagator of the $f$ modes is required
in order to determine the normalization factor of the spinors.
Here we give the details of the formulas that we have quoted in
\Section{sec-spinors} for that purpose. The formulas we present here
are also appropriate for calculating amplitudes of processes involving
the $f$ modes in intermediate states, including possible thermal effects
when required. 

Going back to \Eqs{L0k}{Sigmai}, the propagator is the inverse of 
$\lslash{k} - \Sigma$, or
\beq
S^{-1}_{F0}(k) = \lslash{k} + \mu_L\lslash{u}L + \mu_R\lslash{u}R
- mL - m^\ast R\,,
\eeq
which we write explicitly in the Weyl representation,
\beq
S^{-1}_{F0}(k) = \left(\begin{array}{cc}
    -m^\ast & \omega + \mu_L + \vec\sigma\cdot\vec\kappa\\
    \omega + \mu_R - \vec\sigma\cdot\vec\kappa & -m
    \end{array}\right)\,.
\eeq
Using
\beqa
P_{+} + P_{-} & = & 1\,,\nonumber\\
P_{+} - P_{-} & = & \vec\sigma\cdot\hat\kappa\,,
\eeqa
where $P_s$ are the matrices defined in \Eq{Ps}, we can write
the propagator in the form
\beq
S^{-1}_{F0}(k) = \sum_s S^{-1}_s P_s\,,
\eeq
where
\beq
S^{-1}_s =  \left(\begin{array}{cc}
  -m^\ast & \omega + \mu_L + s\kappa\\
  \omega + \mu_R - s\kappa & -m
\end{array}\right)\,.
\eeq
It then follows immediately that
\beq
\label{SF0exact}
S_{F0} = \sum_s S_s P_s\,,
\eeq
where
\beq
S_s  \equiv \frac{N_s}{d_s}\,,
\eeq
with
\beqa
\label{dN}
d_s(\omega) & = &
(\omega + \mu_L + s\kappa)(\omega + \mu_R - s\kappa) - |m|^2\,,
\nonumber\\[12pt]
N_s(\omega) & = & \left(\begin{array}{cc}
  m & \omega + \mu_L + s\kappa\\
  \omega + \mu_R - s\kappa & m^\ast
\end{array}\right)\,.
\eeqa
It follows [see \Eq{dreq}] that the dispersion relations are the solutions to
\beq
d_s = 0\,,
\eeq
and indeed
\beq
d_s = (\omega - \omegap_s)(\omega - \omegam_s)\,.
\eeq
Therefore,
\beqa
S_s & = & \frac{N_s(\omega)}{(\omega - \omegap_s)(\omega - \omegam_s)}
\nonumber\\
& = & \frac{1}{\omegap_s - \omegam_s}
\left[
\frac{N_s(\omega)}{\omega - \omegap_s} - \frac{N_s(\omega)}{\omega - \omegam_s}
\right]\,.
\eeqa
Following the standard (Feynman) prescription, the poles of the
propagator are to be shifted by replacing\footnote{This is required
for consistency with the one-particle approximation term as well.}
\beq
\omega_s^{(\pm)} \rightarrow \omega_s^{(\pm)} \mp i\epsilon\,.
\eeq
Equivalently, in terms of the particle and antiparticle dispersion relations
$\omega_s$ and $\bar\omega_s$,
\beqa
\omega_s & \rightarrow & \omega_s - i\epsilon\nonumber\\
\bar\omega_s & \rightarrow & \bar\omega_s - i\epsilon\,.
\eeqa

Thus, for $\omega$ near one of the $\omegap_s$ poles,
the propagator $S_{F0}$ given in \Eq{SF0exact} reduces to
\beq
\label{1pomegap}
\left.S_{F0}\right|_{\omega\rightarrow\omegap_s} = 
\frac{N_s(\omegap_s)}
{(\omegap_s - \omegam_s)(\omega - \omegap_s + i\epsilon)} P_s\,.
\eeq
The wavefunction normalization factor $z_s$ in \Eq{ubaru}
is determined by the condition that this coincides with the expression
for $(S_{F0})_{1p}$ given in \Eq{SF01p}. Using \Eqs{DDprime}{ubaru}
we arrive at \Eq{zs}.
Similarly, for the $v$ spinors, near the $\omegam_s$ pole,
\beqa
\label{1pomegam}
\left.S_{F0}\right|_{\omega\rightarrow\omegam_s} & =  &
-\frac{N_s(\omegam_s)}{(\omegap_s - \omegam_s)
(\omega - \omegam_s - i\epsilon)} P_s\nonumber\\
& = & -\frac{N_s(-\bar\omega_s)}
{(\omegap_s - \omegam_s)(\omega + \bar\omega_s - i\epsilon)} P_s\,.
\eeqa
This must be compared with \Eq{SF01barp}. Using \Eq{vbarv} and \Eq{dN}
we arrive at \Eq{zsbar}.

The propagator can be written in covariant form, analogous to the
$u\bar u$ and $v\bar v$ projection operators, which is more convenient
for practical calculations.
Firstly, introducing the off-shell version of $D_{L,R}$,
\beqa
E_{Ls}(\omega,\vkappa) & = & \omega + s\kappa + \mu_L\,,\nonumber\\
E_{Rs}(\omega,\vkappa) & = & \omega - s\kappa + \mu_R\,.
\eeqa
we can write
\beq
\label{Nscovdef}
N_s = (mR + m^\ast L + E_{Ls}\lslash{u} L + E_{Rs}\lslash{u} R)\,.
\eeq
Notice that
\beqa
D_{L,R}(\vkappa) & = & E_{L,R}(\omega_s,\vkappa)\,,\nonumber\\
\bar D_{L,R}(\vkappa) & = & -E_{L,R}(-\bar\omega_s,\vkappa)\,.
\eeqa
Thus,
\beq
S_{F0} = \sum_s \frac{N_s P_s}
{(\omega - \omegap_s + i\epsilon)(\omega - \omegam_s - i\epsilon)}\,,
\eeq
or in terms of the particle and anti-particle dispersion relations
$\omega_s$ and $\bar\omega_s$,
\beqa
\label{SF0final}
S_{F0} & = & \sum_s \frac{N_s P_s}
{(\omega - \omega_s + i\epsilon)(\omega + \bar\omega_s - i\epsilon)}\nonumber\\
& = & \sum_s \frac{1}{\omega_s + \bar\omega_s}
\left[\frac{N_s P_s}{\omega - \omega_s + i\epsilon} -
\frac{N_s P_s}{\omega + \bar\omega_s - i\epsilon}\right]\,,
\eeqa
with $P_s$ and $N_s$ given in \Eqs{Pscovdef}{Nscovdef}, respectively.

The thermal propagator can be constructed
by the usual rules. Denoting the propagator of the $f$ fermion given 
in \Eq{SF0final} by $S^{(f)}_{F0}(k)$, we have, for example,
\beq
\label{SF11}
S^{(f)}_{F11}(q) = S^{(f)}_{F0}(q) - 
\left(S^{(f)}_{F0}(q) - \bar S^{(f)}_{F0}(q)\right)
\eta_f(q)\,,
\eeq
\beq
\label{SF12}
S^{(f)}_{F12}(q) = -\left(S^{(f)}_{F0}(q) - \bar S^{(f)}_{F0}(q)\right)
[\eta_f(q) - \theta(-q\cdot u)]\,,
\eeq
and so on. The factor $\eta_f$ is the usual one, but the chemical potentials
do not appear in the distribution functions, that is
\beq
\eta_f(q) = \theta(q\cdot u)n_F(x_f) + \theta(-q\cdot u)n_F(-x_f)\,,
\eeq
where
\beq
\label{nF}
n_F(x) = \frac{1}{e^x + 1}\,,
\eeq
with
\beq
\label{xf}
x_f = \beta q\cdot u\,.
\eeq

Although the chemical potentials do not appear explicitly
in \Eqs{nF}{xf}, their presence is taken into account
through the dispersion relations, wavefunctions and propagators
of the modes.  Calculations of rates of processes including
the fermion modes, would involve the use of the propagators defined above.
Thus, a calculation of the transition probability for any process
involving the fermion modes would depend on the fermion
distribution functions (e.g., the Pauli blocking factors of the fermion
modes).

\section{Examples}
\label{sec-scatt}

As examples that illustrate the applicability
of the formulas we have obtained and their usefulness,
we will consider the calculation of the rates for  
processses of the type
\beq
f + \chi \rightarrow f + \chi\,,
\eeq
and we will present the results and details
specifically for the case that $\chi$ is a heavy fermion, in the sense that
$\kappa \ll m_\chi$, and essentially at rest.
Due to the fact that the momentum dependence of the
dispersion relations and spinors of the $f$ fermions
are not standard, the kinematics involved in the calculations of the rates
for these processes also differ from the standard case of a free $f$.
In principle, a complete calculation of such
rates would involve their thermal averages, taking into
account the fermion distribution functions.
In this sense, the calculations we present below are useful for 
considering possible application of the results to situations
of phenomenological interest (for example, in the context of DM physics),
and the results can serve as benchmark guides for further exploration
of the type of model we are considering, in specific realistic settings.

A simple possible type of $f\chi$ interaction that we can consider is
(a vector-like interaction)
\beq
L_{f\chi} = g_{LL}(\bar f_L\gamma_\mu f_L)(\bar \chi_L\gamma^\mu \chi_L)\,,
\eeq
which in terms of the primed $f$ field becomes
\beq
L_{f\chi} = g_{LL}(\bar f^\prime_L\gamma_\mu f^\prime_L)
(\bar \chi_L\gamma^\mu \chi_L)\,,
\eeq
or variations of this; for example,
\beq
L_{f\chi} = g_{RL}(\bar f^\prime_L\gamma_\mu f^\prime_L)
(\bar \chi_R\gamma^\mu \chi_R)\,,
\eeq
and similar ones with $(L\leftrightarrow R)$.

As a side note we mention that although it could be thought that
the simplest interaction we can consider are scalar-type interactions
of the form
\beq
L_{f\chi} = g_L(\bar f_R f_L)(\bar \chi_R \chi_L) + H.c.\,,
\eeq
this is actually not consistent. This interaction does not respect the
global symmetries of the model we are considering (Model-I),
corresponding to the global charges
\beqa
Q_1(f_L) = Q_1(f_R) = 1,& \qquad & Q_1(\phi) = 0\,,\nonumber\\
Q_2(\phi) = Q_2(f_R) = 1, & \qquad & Q_2(f_L) = 0\,.
\eeqa
In other words, in terms of the primed fields this interaction term
would become
\beq
L_{f\chi} = g_L e^{-i(\mu_L - \mu_R)t}
(\bar f^\prime_R f^\prime_L)(\bar \chi_R \chi_L) + h.c.\,,
\eeq
which introduces a time-dependent coupling.
Therefore we stick to one of the vector interactions, which are
consistent with the symmetries of Model-I.

For definiteness we consider the $g_{LL}$ interaction. Dropping
the prime superscript on $f$,
\beq
L_{f\chi} = g_{LL}(\bar f_L\gamma_\mu f_L)(\bar \chi_L\gamma^\mu \chi_L)\,,
\eeq
with the understanding that $f$ now stands for the fermion field
of the fermion modes in the BE condensate.
We want to calculate the rates for
\beq
f(k,s) + \chi(p,\lambda) \rightarrow 
f(k^\prime,s^\prime) + \chi(p^\prime,\lambda^\prime)\,,
\eeq
using obvious labels for the kinematic variables.
We write $k^\mu$ and $p^\mu$ in the form
\beqa
\label{kpparametrization}
k^\mu & = & (\omega_s,\vec\kappa)\,,\nonumber\\
p^\mu & = & (E,\vec P)\,,
\eeqa
and similarly
\beqa
k^{\prime\,\mu} & = & (\omega^\prime_{s^\prime},\vec\kappa^\prime)\,,\nonumber\\
p^{\prime\,\mu} & = & (E^\prime,\vec P^\prime)\,,
\eeqa

The amplitude is
\beq
M = g_{LL}(\bar u_f(\vkappa^\prime,s^\prime)\gamma_\mu Lu_f(\vkappa,s))
(\bar u_\chi(\vecP^\prime,\lambda^\prime)
\gamma^\mu Lu_\chi(\vecP,\lambda))\,.
\eeq
To calculate the probability, we average over $\lambda$ (unpolarized
$\chi$) sum over $\lambda^\prime$, but keep $s$ and $s^\prime$ since
in the end we have to multiply by the $f$ distribution functions,
which depend on the helicity of $f$. Thus,
\beq
\overline{|M|^2} =
g^2_{LL}\left(\frac{1}{2}
\left(\vphantom{\frac{1}{2}}
\Tr\gamma_\mu L\rho_f\gamma_\nu L\rho^\prime_f\right)
\Tr\gamma^\mu L\rho_\chi\gamma^\nu L\rho^\prime_\chi\right)\,.
\eeq
For $\chi$,
\beq
\rho_\chi = \lslash{p} + m_\chi\,,
\eeq
and similarly for $\rho^\prime_\chi$. For $f$, 
\beq
\rho_f = u_f(\vkappa,s)\bar u_f(\vkappa,s)\,,
\eeq
and similarly for $\rho^\prime_f$,
where $u_f(k,s)$ is the spinor for the fermions in the BE Model-I.
Here we will use the expression we have obtained for the projection
matrices $\bar u_f u_f$, given in \Eq{uprojcov} and summarized in
\Appendix{sec-appendixA}, namely,
\beq
\label{rhofcov}
\rho_f = \frac{z^2_s}{2}
(mR + m^\ast L + D_{Ls}\lslash{u} L + D_{Rs}\lslash{u} R)
(1 - s\gamma^5\lslash{u}\lslash{n})\,.
\eeq
The formulas for $\rho^\prime_f$ are obtained by
replacing $\vkappa\rightarrow\vkappa^\prime$ and $s\rightarrow s^\prime$
in the corresponding formulas. 

\subsection{Evaluation of the traces}

For $\chi$, following standard procedure,
\beqa
\label{L}
L^{\mu\nu} & \equiv &
\frac{1}{2}\Tr\gamma^\mu L\rho_\chi\gamma^\nu L\rho^\prime_\chi\nonumber\\
& = &
[p^\mu p^{\prime\,\nu} + p^{\prime\,\mu} p^\nu - 
p\cdot p^\prime g^{\mu\nu} + 
i\epsilon^{\mu\nu\alpha\beta}p_\alpha p^\prime_\beta] \,.
\eeqa

\subsubsection{$f$ part}

For the $f$ part, we can simplify the calculation by noticing that
\beq
\Tr\gamma_\mu L\rho_f\gamma_\nu L\rho^\prime_f =
\Tr\gamma_\mu (L\rho_f R)\gamma_\nu(L\rho^\prime_f R)\,,
\eeq
and from \Eq{rhofcov}\footnote{%
It is instructive to notice that in the standard case,
\begin{displaymath}
L(\lslash{k} + m)(1 + s\gamma^5\lslash{\ell})R =
L(\lslash{k} - sm\lslash{\ell}) = 
L(\omega - s\kappa)(\lslash{u} - s\lslash{n})\,,
\end{displaymath}
which agrees with what we get if we take the
$\mu_{L,R}\rightarrow 0$ limit (and $m = m^\ast$) in the following
expression for $L\rho_f R$.
In the second equality we have used $k^\mu = \omega u^\mu + \kappa n^\mu$
and $m\ell^\mu = \kappa u^\mu + \omega n^\mu$, and therefore
\begin{displaymath}
k^\mu - sm\ell^\mu = (\omega - s\kappa)u^\mu + (\kappa - s\omega)n^\mu =
(\omega - s\kappa)u^\mu + (s\kappa - \omega)sn^\mu =
(\omega - s\kappa)(u^\mu - sn^\mu)\,. 
\end{displaymath}
}
\beq
L\rho_f R = \frac{z^2_s}{2}
D_{Rs} L\lslash{u}(1 - s\lslash{u}\lslash{n})R =
\frac{z^2_s}{2}
D_{Rs}L(\lslash{u} - s\lslash{n})\,,
\eeq
and similarly for $\rho^\prime_f$,
\beq
L\rho^\prime_f R = \frac{z^{\prime\,2}_{s^\prime}}{2}
D^\prime_{Rs^\prime} L(\lslash{u} - s^\prime\lslash{n}^\prime)\,.
\eeq
Therefore,
\beq
\Tr\gamma_\mu L\rho_f\gamma_\nu L\rho^\prime_f =
\frac{z^2_s}{2}\frac{z^{\prime\,2}_{s^\prime}}{2}
D_{Rs}D^\prime_{Rs^\prime} (2F_{\mu\nu})\,,
\eeq
where
\beq
F_{\mu\nu} \equiv 
\frac{1}{2}\Tr\left\{R\gamma_\mu (\lslash{u} - s\lslash{n})\gamma_\nu
(\lslash{u} - s^\prime\lslash{n}^\prime)\right\}\,,
\eeq
which can be evaluated by using the same formulas for the traces
that we have used in the $\chi$ part. The result then is the same
that we have obtained above for $L_{\mu\nu}$, with the replacements,
\beqa
p_\mu & \rightarrow & q_\mu \equiv u_\mu - sn_\mu\,,\nonumber\\
p^{\prime}_\mu & \rightarrow & q^{\prime}_\mu \equiv
u_\mu - s^\prime n^{\prime}_\mu\,.
\eeqa
Thus,
\beq
\label{F}
F_{\mu\nu} = [q_\mu q^{\prime}_\nu + q^{\prime}_\mu q_\nu - 
q\cdot q^\prime g_{\mu\nu} + 
i\epsilon_{\mu\nu\alpha\beta}q^\alpha q^{\prime\,\beta}] \,,
\eeq
with $q$ and $q^\prime$ defined above.

The amplitude squared is then given by
\beq
\overline{|M|^2} =
\frac{1}{2}g^2_{LL}z^2_s z^{\prime\,2}_{s^\prime}
D_{Rs}D^\prime_{Rs^\prime} L^{\mu\nu}F_{\mu\nu}\,,
\eeq
with $L^{\mu\nu}$ and $F^{\mu\nu}$ given in \Eqs{L}{F}.
By straightforward algebra, it follows that
\beq
L^{\mu\nu} F_{\mu\nu} = 4(p\cdot q)(p^\prime\cdot q^\prime)\,,
\eeq
with
\beqa
p\cdot q & = & E + s\vec P\cdot\hat\kappa\,,\nonumber\\
p^\prime\cdot q^\prime & = & E^\prime +
s^\prime\vec P^\prime\cdot\hat\kappa^\prime\,.
\eeqa

\subsection{
  $f + \chi\rightarrow f +\chi$
  with $\chi$ at rest (heavy $\chi$)
}
\label{sec-fN}

We now consider in detail the case that $\chi$
is initially at rest, and heavy, i.e., with a mass
$m_\chi$ such that
\beq
\label{kappallmX}
\kappa \ll m_\chi\,.
\eeq
In a physical setting, $\chi$ can stand for a nucleon, with the process
taking place with $\kappa$ much less than the nucleon mass, or some other
particle in the medium, but which is not part of the BE condensate background.
Because the $f$ mode is an excitation of the BE condensate,
rather than a particle that travels from far away and scatters
from a target, the quantity of interest is the process rate,
typically denoted by $v\sigma$. Denoting by $n_\chi$ the number
density of $\chi$ particles, the total transition probability
per second is given by $n_\chi v\sigma$.
Our starting point to determine $v\sigma$ is,
\beq
\label{sigmadef}
v d\sigma = \frac{1}{4E\omega_s} \overline{|M|^2} dI\,,
\eeq
where $dI$ is the phase space integral
\beq
dI \equiv (2\pi)^4\delta^{(4)}(p^\prime + k^\prime - p - k)
\frac{d^3 P^\prime}{(2\pi)^3 2E^\prime}
\frac{d^3 \kappa^\prime}{(2\pi)^3 2\omega^\prime_{s^\prime}}\,.
\eeq

As can be seen from \Eq{drmodeliphysical}, the dispersion
relations depend on the helicity and have a non-standard momentum dependence.
A particular consequence is the fact that there is a region of $\kappa$ in which
the dispersion relation of one of the modes
has negative group velocity. Specifically this occurs for $s = +1$
if $\mu_{-} > 0$, or $s = -1$  if $\mu_{-} < 0$. Considering for definiteness
the case $\mu_{-} > 0$, then $\omega_{+}$ has negative group velocity
for $\kappa < \mu_{-}$, and in fact becomes zero at $\kappa = \mu_{-}$.
This is reminiscent of the Van Hove singularity of the density of
states in Condensed Matter systems\cite{vanhovesing,vanhovenature}.
For definiteness, in the remainder of this work we adopt the convention that
\beq
\label{mumconventon}
\mu_{-} > 0\,.
\eeq
The relevant point for us is that
\beq
\label{vgzero}
\left.\frac{d\omega_{+}}{d\kappa}\right|_{\kappa = \mu_{-}} = 0\,.
\eeq
The opposite convention $\mu_{-} < 0$
can be treated similarly with corresponding interchange of the $s$ helicity
states.

In general, $p^\mu$ is given as in \Eq{kpparametrization},
here we consider the case that $\chi$ is
\begin{enumerate}
  \item initially at rest,
    \beq
    p^\mu = (m_\chi,\vec 0) \qquad \Rightarrow \qquad p\cdot q = m_\chi\,,
    \eeq
    \item heavy,
      \beq
	  \label{largemN}
      \kappa,\omega_s(\kappa) \ll m_\chi\,.
      \eeq
      Energy-momentum conservation implies that 
      \beq
	  \label{EprimelargemN}
      E^\prime \sim m_\chi\,,
      \eeq
      so that we can set
      \beq
      p^\prime\cdot q^\prime \rightarrow m_\chi\,.
      \eeq
\end{enumerate}
Therefore, in the case we are considering,
\beq
L^{\mu\nu} F_{\mu\nu} = 4m^2_\chi\,,
\eeq
and whence
\beq
\label{Msquared}
\overline{|M|^2} =
2 m^2_\chi g^2_{LL}z^2_s z^{\prime\,2}_{s^\prime}
D_{Rs}D^\prime_{Rs^\prime}
\eeq

Regarding the phase-space integral, doing the integral over $P^\prime$,
\beq
dI = \frac{1}{4\pi^2}\frac{1}{2E^\prime}
\delta(E^\prime + \omega^\prime_{s^\prime} - m_\chi - \omega_s)
\frac{d^3 \kappa^\prime}{2\omega^\prime_{s^\prime}}\,,
\eeq
where
\beq
\label{Eprimekappaprime}
E^\prime = \sqrt{|\vec\kappa - \vec\kappa^\prime|^2 + m^2_\chi}
\simeq \frac{|\vec\kappa - \vec\kappa^\prime|^2}{2m_\chi} + m_\chi\,.
\eeq
Therefore,
\beqa
dI & = & \frac{1}{4\pi^2}\frac{1}{2E^\prime}
\frac{\kappa^{\prime\,2}}{2\omega^\prime_{s^\prime}}
\frac{1}{\left|\frac{dE^\prime}{d\kappa^\prime} +
\frac{d\omega^\prime_{s^\prime}}{d\kappa^\prime}\right|}
d\Omega^\prime\,,
\eeqa
where $\kappa^\prime$ must satisfy
\beq
\label{energycons}
\omega^\prime_{s^\prime} + E^\prime = \omega_s + m_\chi\,,
\eeq
or explicitly,
\beq
\label{energyconskappap}
\left[(\kappa^\prime - s^\prime\mu_{-})^2 + |m|^2\right]^{1/2} +
\frac{|\vec\kappa - \vec\kappa^\prime|^2}{2m_\chi} =
\left[(\kappa - s\mu_{-})^2 + |m|^2\right]^{1/2}\,.
\eeq
Defining
\beq
x = \cos\theta^\prime \equiv \hat\kappa\cdot\hat\kappa^\prime\,,
\eeq
we then have
\beq
\label{dI}
dI = \frac{1}{8\pi}
\frac{\kappa^{\prime\,2}}{E^\prime\omega^\prime_{s^\prime}}
\frac{1}{\left|\frac{\kappa^\prime - \kappa x}{m_\chi} +
\frac{d\omega^\prime_{s^\prime}}{d\kappa^\prime}\right|}dx\,.
\eeq

The expression for $v\sigma$ is then
\beq
\label{sigmatotalexpr}
v d\sigma = \frac{1}{4m_\chi\omega_s}\overline{|M|^2} dI\,,
\eeq
where $\overline{|M|^2}$ is given in \Eq{Msquared}.
Given $\kappa$, $s$ and $s^\prime$,
then $\kappa^\prime$ is given by the solution of \Eq{energyconskappap}.
If there are several solutions, then there is a rate for
each such $\kappa^\prime$. The rate has a particular
$\kappa$ dependence. In particular, \Eq{dI} reveals that
if the solution for $\kappa^\prime$ is near the point
$\kappa^\prime = \mu_{-}$, at which
$\frac{d\omega^\prime_{s^\prime}}{d\kappa^\prime}$ is zero,
then the term $\frac{\kappa^\prime - \kappa x}{m_\chi}$ must be
taken into account, and the recoil energy of the $\chi$
in \Eq{energyconskappap} cannot be dropped when determining $\kappa^\prime$.
But in the rest of the factors that appear in the expression for $v\sigma$
we can use the large $m_\chi$ limit and set $E^\prime \rightarrow m_\chi$
and $\omega^\prime_{s^\prime} = \omega_{s}$. We then obtain
\beq
dI = \frac{1}{8\pi}
\frac{\kappa^{\prime\,2}}{m_\chi\omega_{s}}
\frac{1}{\left|\frac{\kappa^\prime - \kappa x}{m_\chi} +
  \frac{d\omega^\prime_{s^\prime}}{d\kappa^\prime}\right|}dx\,,
\eeq
and in $|M|^2$ we can put
\beqa
z^\prime_{s^\prime} & = & z_s\,,\nonumber\\
D^\prime_{Rs^\prime} & = &
\omega_s - s^\prime\kappa^\prime + \mu_R\,,\nonumber\\
& = & \omega_{0s} - s^\prime\kappa^\prime + \mu_{-}\,.
\eeqa
Putting it all together in \Eq{sigmatotalexpr},
\beq
\label{sigmatotal}
v d\sigma = \frac{g^2_{LL}}{16\pi}
\frac{\kappa^{\prime\,2}}{\omega^2_{0s}}
(\omega_{0s} - s\kappa + \mu_{-})(\omega_{0s} - s^\prime\kappa^\prime + \mu_{-})
\frac{1}{\left|\frac{\kappa^\prime - \kappa x}{m_\chi} +
\frac{\kappa^\prime - s^\prime\mu_{-}}{\omega^\prime_{0s^\prime}}\right|}dx\,,
\eeq
where $\omega^\prime_{0s^\prime}$ is given by the expression in \Eq{omega0s},
with the replacements $\kappa,s \rightarrow \kappa^\prime,s^\prime$

We now consider the solutions of \Eq{energyconskappap}, separately one by one.
\begin{figure}
\begin{center}
\epsfig{file=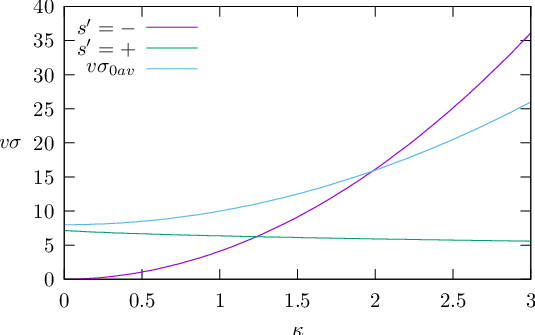,bbllx=154,bblly=294,bburx=409,bbury=456}
\end{center}
\caption[] {
  Plot of $v\sigma$ for $s = -$ and $s^\prime$ as indicated (Case-1), for
  $\mu_{-} = 1, \mu_{+} = 1.5, |m| = 2$. For reference, also plotted is
  $v\sigma_{0av}$, which is calculated by setting
  $\mu_{\pm} = 0$, and then summing over $s^\prime$ and
  averaging over $s$.
  \label{fig1}
}
\end{figure}
\begin{figure}
\begin{center}
\epsfig{file=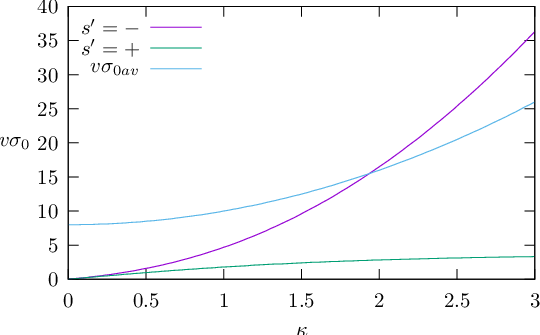,bbllx=152,bblly=294,bburx=409,bbury=456}
\end{center}
\caption[] {
  Plot of $v\sigma$ for $s = -$ and $s^\prime$ as indicated, for
  $\mu_{\pm} = 0, |m| = 2$. For reference, also plotted is
  $v\sigma_{0av}$, which calculated by setting
  $\mu_{\pm} = 0$ as above, and then summing over $s^\prime$
  and averaging over $s$.
  \label{fig2}
}
\end{figure}
\begin{figure}
\begin{center}
\epsfig{file=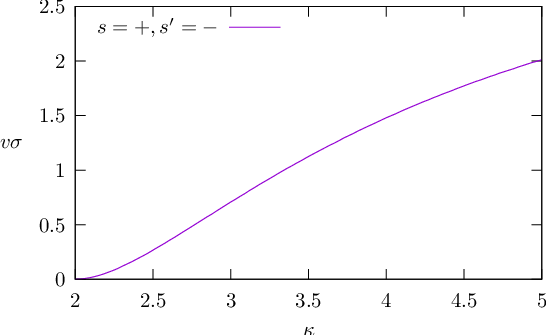,bbllx=150,bblly=294,bburx=409,bbury=456}
\end{center}
\caption[] {
  Plot of $v\sigma$ for $s = +$ and $s^\prime = -$,
  $\mu_{-} = 1, \mu_{+} = 1.5, |m| = 2$. As discussed in the text (Case-2),
  the process can take place provided $\kappa > 2\mu_{-}$.
  \label{fig3}
}
\end{figure}
\begin{figure}
\begin{center}
\epsfig{file=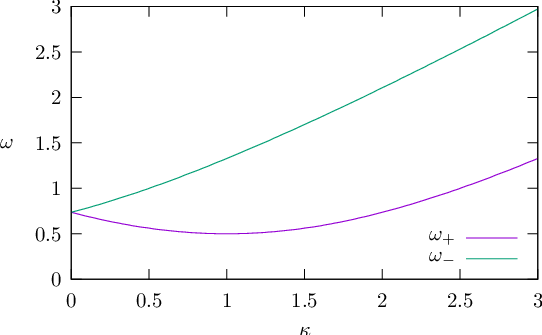,bbllx=149,bblly=294,bburx=409,bbury=456}
\end{center}
\caption[] {
  Plot of dispersion relations, for the sample values
  $\mu_{-} = 1, \mu_{+} = 1.5, |m| = 2$. 
  The lower curve, $\omega_{+}$, has a minimum at $\kappa = \mu_{-}$. Around
  that point, there can be two values of $\kappa$ for a given
  value of $\omega$, which is the reason why there are two possible
  solutions for $\kappa^\prime$ discussed in Case 3 in the text.
  \label{fig4}
}
\end{figure}
\begin{description}
\item[Case 1]  Consider $s = -$. Ignoring the $\chi$ recoil energy term
  in \Eq{energyconskappap}, the equation for $\kappa^\prime$ becomes
  \beq
  \label{kappaprimeeqex1}
  |\kappa^\prime - s^\prime\mu_{-}| = |\kappa + \mu_{-}|\,,
  \eeq
  whose only solution is
  \beq
  \label{kappapexamplecase1}
  \kappa^\prime = \kappa + \mu_{-} + s^\prime \mu_{-}\,.
  \eeq
  In either case, it follows that
  \beqa
  \kappa^\prime - s^\prime\mu_{-} & = & \kappa + \mu_{-}\,,\nonumber\\
  \omega^\prime_{0s^\prime} & = & \omega_{0-}\,,
  \eeqa
  and therefore
  \beq  
  \frac{\kappa^\prime - s^\prime\mu_{-}}{\omega^\prime_{0s^\prime}} =
  \frac{\kappa + \mu_{-}}{\omega_{0-}}\,.
  \eeq
  This implies that keeping the term $(\kappa^\prime - \kappa x)/m_\chi$ in the
  denominator of \Eq{sigmatotal}, as well as the $\chi$ recoil energy
  in \Eq{energyconskappap} would give insignificant corrections
  to the solution given in \Eq{kappapexamplecase1}.
  By straightforward algebra, the following explicit expressions for
  $v\sigma$ follow from \Eq{sigmatotal},
  \beqa
  \label{sigmatotalex1}
  v\sigma & = & \frac{g^2_{LL}}{8\pi}
  \frac{(\kappa + \mu_{-} + s^\prime\mu_{-})^2}{(\kappa + \mu_{-})\omega_{0-}}
  \left[\omega_{0-} + \kappa + \mu_{-}\right]
  \left[\omega_{0-} - s^\prime(\kappa + \mu_{-})\right]\nonumber\\[12pt]
  & = & \frac{g^2_{LL}}{8\pi}\left\{
  \frac{1}{(\kappa + \mu_{-})\omega_{0-}}
  \right\}
  \left\{\begin{array}{ll}
  (\kappa + 2\mu_{-})^2|m|^2 & (s^\prime = +)\\[12pt]
  \kappa^2
  \left\{\sqrt{(\kappa + \mu_{-})^2 + |m|^2} + \mu_{-} + \kappa\right\}^2
  & (s^\prime = -)
  \end{array}\right.
  \eeqa
  Sample plots of $v\sigma$ given in \Eq{sigmatotalex1} are given in \Fig{fig1}.
  For comparison and reference, \Fig{fig2} shows the plot
  of $v\sigma$ calculated by setting $\mu_{\pm} = 0$.
\item[Case 2] Consider $s = +$, $s^\prime = -$.
  Again ignoring the $\chi$ recoil energy in \Eq{energyconskappap},
  the equation for $\kappa^\prime$ becomes
  \beq
  \kappa^\prime = -\mu_{-} \pm |\kappa - \mu_{-}|\,.
  \eeq
  Only the solution with the positive sign is possible.
  Then a solution exists (the process is possible) provided
  \beq
  \label{kappapexamplecase2cond}
  \kappa > 2\mu_{-}\,,
  \eeq
  so that
  \beq
  \label{kappapexamplecase2sol}
  \kappa^\prime = \kappa - 2\mu_{-} \ge 0\,.
  \eeq
  Furthermore,
  \beq  
  \frac{\kappa^\prime - s^\prime\mu_{-}}{\omega^\prime_{0s^\prime}} =
  \frac{\kappa - \mu_{-}}{\omega_{0+}} \ge \frac{\mu_{-}}{\omega_{0+}}\,,
  \eeq
  which implies, as in Case-1 above,  that keeping the term
  $(\kappa^\prime - \kappa x)/m_\chi$ in the denominator of \Eq{sigmatotal},
  as well as the $\chi$ recoil energy in \Eq{energyconskappap} would give
  insignificant corrections to the solution given in
  \Eq{kappapexamplecase2sol}.
  In this case the following explicit form for $v\sigma$ follows from
  \Eq{sigmatotal},
  \beq
  v\sigma = \frac{g^2_{LL}}{8\pi}
  \frac{(\kappa - 2\mu_{-})^2}{(\kappa - \mu_{-})\omega_{0+}}
  |m|^2 \qquad
  (\kappa > 2\mu_{-})\,.
  \eeq
  Due to the kinematical restriction $\kappa > 2\mu_{-}$,
  the apparent singularity at $\kappa = \mu_{-}$
  actually lies outside the physical region.   
  Sample plot of $v\sigma$ in this case is shown in \Fig{fig3}.

  \item[Case 3] Consider $s = +$, $s^\prime = +$. In this case it is
  necessary to include the term $(\kappa^\prime - \kappa x)/m_\chi$
  in \Eq{sigmatotal} and take into account the $\chi$ recoil energy
  in \Eq{energyconskappap}. To see this, let us pretend that we can
  neglect them and proceed as in the previous Case-1 and Case-2 above.
  Then \Eq{energyconskappap} gives
  \beq
  \kappa^\prime = \mu_{-} \pm |\kappa - \mu_{-}|\,,
  \eeq
  which can have two solutions for $\kappa^\prime$.
  As a complement to understanding the origin of the two solutions
  for $\kappa^\prime$, the dispersion relations are plotted in \Fig{fig4}.
  For $\omega_{+}$ around its minimum value, the equation
  $\omega_{+}(\kappa) = \omega$ gives two values of $\kappa$. 
  Thus, if $\kappa < \mu_{-}$, both values of $\kappa^\prime$ are allowed,
  \beq
  \kappa^\prime = \left\{
    \begin{array}{ll}2\mu_{-} - \kappa & (+)\\ \kappa & (-)\end{array}
    \right.
  \eeq
  If $\mu_{-} < \kappa < 2\mu_{-}$, again both values of $\kappa^\prime$
  are allowed,
  \beq
    \kappa^\prime = \left\{
    \begin{array}{ll}2\mu_{-} - \kappa & (-)\\ \kappa & (+)\end{array}
    \right.
  \eeq
  If $\kappa > 2\mu_{-}$,
  \beq
  \kappa^\prime = \kappa\,.
  \eeq
  In other words, in summary, the process
  $f_{+}(\vkappa) \rightarrow f_{+}(\vkappa^\prime)$ can occur with
  \begin{description}
    \item[Case 3a]
      \beq
      \kappa^\prime = \kappa \qquad \mbox{(all $\kappa$)}\,,
      \eeq
      \Eq{sigmatotal} can be written in the following explicit form,
      \beq
      \label{sigma3a}
      v\sigma = \frac{g^2_{LL}}{8\pi}
      \frac{\kappa^2}{|\kappa - \mu_{-}|\omega_{0+}}
      \left(\sqrt{(\kappa - \mu_{-})^2 + |m|^2} + \mu_{-} - \kappa\right)^2
      \qquad \mbox{all $\kappa$}\,,
      \eeq
      which literally exposes the singularity at $\kappa = \mu_{-}$.
    \item[Case 3b]
      \beq
      \kappa^\prime = 2\mu_{-} - \kappa \qquad (\kappa < 2\mu_{-})\,.
      \eeq
      Similarly, the result in this case is,
    \beq
    \label{sigma3b}
    v\sigma = \frac{g^2_{LL}}{8\pi}
    \frac{(2\mu_{-} - \kappa)^2 |m|^2}{|\kappa - \mu_{-}|\omega_{0+}}\qquad
    (\kappa^\prime = 2\mu_{-} - \kappa \mbox{ with } \kappa < 2\mu_{-})
    \eeq
    again illustrating the singularity at $\kappa = \mu_{-}$.
  \end{description}
\end{description}

The lesson is that the $(\kappa^\prime - \kappa x)/m_\chi$ term
cannot be neglected, and the $\chi$ recoil energy
in \Eq{energyconskappap} must be taken into account when solving for
$\kappa^\prime$, when $\kappa$ is in the region $\kappa \simeq \mu_{-}$.
The above formulas for $v\sigma$ in Case-3a and 3b can be used
for $\kappa$ away from the singularity $\kappa \simeq \mu_{-}$.

\subsection{Discussion}
\label{sec:scatt-discussion}

It is encouraging that the final formulas for $v\sigma$
that we have obtained in the examples considered above,
look compact and simple.  However, we should
not miss the fact that altogether they describe some
rather peculiar effects. First of all the kinematics differs
from the one in the standard case (with the vacuum dispersion relations),
which can be traced back to the fact that the dispersion
relations depend on the helicity and have a non-standard momentum dependence.

On the opposite side of the case of a heavy $\chi$ that we have considered
above, is the case of an extremely relativistic, or massless $\chi$,
with the initial $f$ being at rest, in symbols,
\beq
\chi(\vec P) + f(\vec\kappa = 0) \rightarrow \chi(\vec P^\prime) +
f(\vec\kappa^\prime)\,,
\eeq
where the $\chi$ particle is extremely relativistic, or massless.
Momentum conservation gives the recoil $f$ momentum
\beq
\vec\kappa^\prime = \vec P - \vec P^\prime\,,
\eeq
and the energy conservation condition requires
\beq
P^\prime +
\sqrt{\left(|\vec P - \vec P^\prime| - s^\prime\mu_{-}\right)^2 + |m|^2} = P +
\sqrt{\mu^2_{-} + |m|^2}\,.
\eeq
We have carried out some preliminary work to calculate the cross
section in this case, which also hints to
peculiar and interesting features in its own right.

\section{Summary and outlook}
\label{sec-conclusions}

In summary, in this work we have considered the determination of
the fermion spinors, the corresponding projection matrices and
propagators, in the models considered in \Rref{ns:fbec} for the propagation
of fermions interacting with a BE condensate. 
We have obtained concise formulas which are useful for calculations of
process rates involving the fermion modes of the BE condensate.
For convenience and future reference, we have summarized in
\Appendix{sec-appendixA} the formulas for the fermion dispersion relations,
spinors, projection operators and propagators that are useful
for reference in future works.

As an application we have considered the calculation
of the scattering rate of the fermion mode $f$
of the BE condensate, with a generic fermion which we denoted by $\chi$,
using commonly used models for the $f\chi$ interactions.
For definiteness, we considered in detail the case in which
$\chi$ is a heavy fermion, initially at rest. Even in this relatively simple
situation the analysis and results illustrate
several peculiarities that must dealt with in this type of calculation
involving the $f$ fermion modes. They are a consequence of the fact that their
dispersion relations are not the standard vacuum
dispersion relations, but actually are helicity-dependent,
and therefore the kinematics differ from the case in which the fermions
propagate in vacuum.

In general terms, the results we have presented here, besides
their value in their own right and serving as a guide for treating
similar models, can be useful in specific contexts and applications,
for example involving neutrino, electron or nucleon propagation
and interactions with dark matter particles.

The work of S. S. is partially supported by 
DGAPA-UNAM (Mexico) PAPIIT Project No. IN105326.

\appendix

\section{Summary}
\label{sec-appendixA}

For convenience we summarize here the main formulas for the
fermion dispersion relations, spinors and propagator.

\subsection{Dispersion relations}

\beqa
\mbox{particle}: \omega_{s}(\vec\kappa) & = &
\left[(\kappa - s\mu_{-})^2 + |m|^2\right]^{1/2} - \mu_{+}\,,\nonumber\\
\mbox{antiparticle}: \bar\omega_{s}(\vec\kappa) & = &
\left[(\kappa - s\mu_{-})^2 + |m|^2\right]^{1/2} + \mu_{+}\,,
\eeqa
with
\beq
\mu_{\pm} \equiv \frac{1}{2}(\mu_R \pm \mu_L)\,.
\eeq

\subsection{Spinors}

\beqa
\mbox{particle}: u_s(\vkappa) & = & z_s\left(
\begin{array}{cc}\sqrt{D_{Ls}}\\[12pt] \alpha^\ast\sqrt{D_{Rs}}
\end{array}\right)\chi_s(\vkappa)\nonumber\\
\mbox{antiparticle}: v_s(\vkappa) & = & \bar z_s\left(
\begin{array}{cc} \sqrt{\bar D_{Ls}}\\[12pt] -\alpha^\ast\sqrt{\bar D_{Rs}}
\end{array}\right)\chi_s(-\vkappa)\,,
\eeqa
where
\beq
(\vec\sigma\cdot\hat\kappa)\chi_s(\vkappa) =
s\chi_s(\vkappa) \qquad (\mbox{helicity spinor})\,,
\eeq
\beqa
D_{Ls}(\vkappa) & = & \omega_s + s\kappa + \mu_L\,,\nonumber\\
D_{Rs}(\vkappa) & = & \omega_s - s\kappa + \mu_R\,,
\eeqa
\beqa
\bar D_{Ls}(\vkappa) & = & \bar\omega_s - s\kappa - \mu_L\,,\nonumber\\
\bar D_{Rs}(\vkappa) & = & \bar\omega_s + s\kappa - \mu_R\,,
\eeqa
\beqa
z_s(\vkappa) & = & \left\{
\frac{\omega_s(\vkappa)}{\sqrt{(\kappa - s\mu_{-})^2 + |m|^2}}
\right\}^{\frac{1}{2}}\,,\nonumber\\
\bar z_s(\vkappa) & = & \left\{
\frac{\bar\omega_s(\vkappa)}{\sqrt{(\kappa - s\mu_{-})^2 + |m|^2}}
\right\}^{\frac{1}{2}}\,,\nonumber\\
\alpha & = & \frac{m}{|m|}\,.
\eeqa

\subsection{Projection matrices}

\beqa
u_{s}\bar u_{s} & = & z^2_s
\left(\begin{array}{cc}
  m & D_{Ls}\\ D_{Rs} & m^\ast
\end{array}\right)P_s(\vkappa)\,,\nonumber\\
v_{s} \bar v_{s} & = & \bar z^2_s
\left(\begin{array}{cc}
  -m & \bar D_{Ls}\\ \bar D_{Rs} & -m^\ast
\end{array}\right)P_s(-\vkappa)\,,
\eeqa
where
\beq
P_s(\vkappa) \equiv \frac{1}{2}(1 + s\vec\sigma\cdot\hat\kappa)\,.
\eeq
An alternative form that may be convenient for some calculations is,
\beqa
u_s \bar u_s & = & \frac{z^2_s}{2}
(mR + m^\ast L + D_{Ls}\lslash{u} L + D_{Rs}\lslash{u} R)
(1 - s\gamma^5\lslash{u}\lslash{n})\,,\nonumber\\
v_s \bar v_s & = & \frac{\bar z^2_s}{2}
(-mR - m^\ast L + \bar D_{Ls}\lslash{u} L + \bar D_{Rs}\lslash{u} R)
(1 + s\gamma^5\lslash{u}\lslash{n})\,,
\eeqa
where
\beqa
u^\mu & = & (1,\vec 0)\,,\nonumber\\
n^\mu & = & (0,\hat\kappa)\,.
\eeqa
In the limit $\mu_{R,L} = 0$ and $m^\ast = m$ these formulas
reduce to the standard vacuum formulas, as it should be,
\beqa
u_s \bar u_s & = & \frac{1}{2}(\lslash{k} + m)(1 + s\gamma^5\lslash{\ell})\,,
\nonumber\\
v_s \bar v_s & = & \frac{1}{2}(\lslash{k} - m)(1 + s\gamma^5\lslash{\ell})\,,
\eeqa
where
\beq
\ell^\mu = \frac{1}{m}(\kappa u^\mu + \omega n^\mu) =
\frac{1}{m}(\kappa,\omega\hat\kappa)\,.
\eeq

\subsection{Propagator}
\beq
S_{F0} = \sum_s S_s P_s(\vkappa)\,,
\eeq
where,
\beqa
S_s & = & 
\frac{N_s}{(\omega - \omega_s + i\epsilon)(\omega + \bar\omega_s - i\epsilon)}
\nonumber\\
& = & \frac{1}{\omega_s + \bar\omega_s}
\left[
  \frac{N_s}{\omega - \omega_s + i\epsilon} -
  \frac{N_s}{\omega + \bar\omega_s - i\epsilon}
\right]\,,
\eeqa
\beq
N_s = \left(\begin{array}{cc}
  m & \omega + \mu_L + s\kappa\\
  \omega + \mu_R - s\kappa & m^\ast
\end{array}\right)\,.
\eeq

An alternative form analogous to $u\bar u$
and $v\bar v$ above, which is more convenient for practical calculations
in particular those involving the thermal propagators, is to use
\beq
\label{Pssummary}
P_s = \frac{1}{2}(1 - s\gamma^5\lslash{u}\lslash{n})\,,
\eeq
and
\beq
\label{Nssummary}
N_s = (mR + m^\ast L + E_{Ls}\lslash{u} L + E_{Rs}\lslash{u} R)\,,
\eeq
where $E_{L,R}$ are the off-shell version of $D_{L,R}$,
\beqa
E_{Ls}(\omega,\vkappa) & = & \omega + s\kappa + \mu_L\,,\nonumber\\
E_{Rs}(\omega,\vkappa) & = & \omega - s\kappa + \mu_R\,.
\eeqa
Notice that
\beqa
D_{L,R} & = & E_{L,R}(\omega_s,\vkappa)\nonumber\\
\bar D_{L,R} & = & -E_{L,R}(-\bar\omega_s,\vkappa)\,.
\eeqa
In summary,
\beqa
S_{F0} & = & \sum_s \frac{N_s P_s}
{(\omega - \omega_s + i\epsilon)(\omega + \bar\omega_s - i\epsilon)}\nonumber\\
& = & \sum_s \frac{1}{\omega_s + \bar\omega_s}
\left[\frac{N_s P_s}{\omega - \omega_s + i\epsilon} -
\frac{N_s P_s}{\omega + \bar\omega_s - i\epsilon}\right]\,,
\eeqa
with $P_s$ and $N_s$ given in \Eqs{Pssummary}{Nssummary}.


\begin{thebibliography}{99}
\bibitem{footnote1}
  See, for example,
  Refs.\ \cite{weldon:phimu}, \cite{kapusta:be}, \cite{haberweldon:be}.
\end{thebibliography}

\end{document}